\documentstyle[aps]{revtex}
\begin{document}
\draft
\title{Cascade of two-dimensional Fulde-Ferrell-Larkin-Ovchinnikov phases with anisotropy}
\author{R. Combescot$^{a}$ and G. Tonini$^{b}$}
\address{$^{a}$Laboratoire de Physique Statistique,
 Ecole Normale Sup\'erieure*,
24 rue Lhomond, 75231 Paris Cedex 05, France}
\address{$^{b}$Dipartimento di Fisica, Universit\`a di Firenze, Firenze, Italy}
\date{Received \today}
\maketitle

\begin{abstract}
For an isotropic two-dimensional system, when the temperature is lowered toward $T=0$, it has been found recently that, for the transition from the normal to the superfluid state in the paramagnetic limit, the order parameter describing the Fulde-Ferrell-Larkin-Ovchinnikov phase has an increasingly complex structure with contributions from an increasing number of wavevectors. This cascade of phase transitions is directly linked to the fact that, due to the rotational invariance, all the wavevectors directions which can enter the order parameter are degenerate. We study how this cascade of phase transitions is modified when one takes into account the anisotropy arising in a real solid state compound. For a simple model of anisotropy with elliptical dispersion relation, we find surprisingly that the cascade of phase transitions is not modified and the degeneracy with respect to the wavevector direction is still present. When we take into account a deviation with respect to this elliptical model, which is treated to first order in perturbation, we find that the degeneracy is lifted and that, basically, a single wavevector is favored right at the transition. However when one enters the superfluid phase, additional wavevectors come in the order parameter and the cascade of transitions is still present, though in a modified form.
\end{abstract}

 
\section{INTRODUCTION}
Experimental progress has stimulated current research in the field of Fulde-Ferrell-Larkin-Ovchinnikov (FFLO) phases \cite{ff,larkov}, with the proposal of several new compounds as displaying these phases. Indeed anomalies in the heavy fermion compound CeCoIn$_5$ have been attributed very recently to FFLO phases \cite{movsh}. This corresponds to a standard 3-dimensional situation. However the quasi 2-dimensional case is also of very high interest \cite{shima1} since this is the geometry where one can hope to find compounds with very high critical fields and there have been also recent claims of experimental observations \cite{singleton,tanatar}. In practice this corresponds to compounds made of conducting planes, with very small coupling between these planes. Many organic materials and also high $T_c$ superconductors fall in this class. In such cases for a field strictly parallel to the planes, the orbital currents can not flow perpendicular to the planes. Hence they do not come into play for the determination of the critical field, which is only controlled by the paramagnetic effect. This is just the dominant one which is under study in the FFLO phases. Naturally in practice the field is not exactly parallel to the planes. This brings additional structure due to the appearance of orbital currents and related vortex structures \cite{shra,khr,hb,klein,ym}. Naturally it is important to control first the pure FFLO situation in order to be able to disentangle this quite complex situation.

On the other hand the theoretical situation has been shown recently to be more complex than expected at low temperature. This is in contrast with the case of higher temperature where, below the tricritical point, the order parameter is believed to be a simple cosine $\Delta ( {\bf  
r}) \sim \cos({\bf  q}. {\bf  r})$ at the transition from the normal to the FFLO phase. In the isotropic case, corresponding to a circular Fermi line (that is a cylindrical Fermi surface), the direction of the ${\bf q}$ vector is degenerate. However it was first realized \cite{shima2} by Shimahara that this degeneracy can be used to lower the free energy and that the most favorable state contains more than a single cosine in its order parameter when the temperature is lowered. Quite recently this investigation has been pushed to lower temperature \cite{rccm}, down to zero temperature, and it was discovered that in this regime one has a cascade of phase transitions, with the order parameter of each of the FFLO phases being made of the superposition of an increasing number of cosines when the temperature is lowered.

Once this quite peculiar behaviour of the isotropic case is recognized, it is of interest to investigate what is left of it in real compounds. Indeed no actual solid state compound is exactly isotropic, although in many cases this may be a rather good approximation. In particular the Fermi lines corresponding to the contribution of the $CuO_2$ planes in high $T_c$ compounds are actually fairly circular (with the center of the circle being at ($\pm \pi ,\pm\pi $) in the Brillouin zone), which is somewhat surprising. It seems clear at first that anisotropy will lift the degeneracy with respect to the ${\bf q}$ vector found in the isotropic case, except for some trivial degeneracies occuring because of remaining discrete symmetries. Hence one expects that the transition from the normal state to the FFLO phase will occur for a well defined wavevector ${\bf q}_0$ (and its opposite if parity is conserved), leading again to a $\Delta ( {\bf  r}) \sim \cos({\bf  q}_0. {\bf  r})$ order parameter. This is indeed what we will find in the general situation. Technically this results from the analysis of the second order term in a Ginzburg-Landau expansion. However it is also clear that wavevectors in the vicinity of  ${\bf q}_0$ are nearly as favored for the order parameter as ${\bf q}_0$ itself. As soon as one goes into the FFLO superfluid phase (for example by lowering the temperature), they will become a possible choice. Moreover entering the FFLO phase produces a growth of the fourth order term in the Ginzburg-Landau expansion. As shown by Larkin and Ovchinnikov, this is this term which produces a coupling between the various possible wavevectors. In our specific case, we know from the analysis of the isotropic case that it favors the existence of nearby wavevectors and is ultimately responsible for the cascade found in this isotropic situation. Hence we may rightly expect to see the cascade appear when we penetrate into the FFLO phase, although in a milder form since wavevectors with directions far away from ${\bf q}_0$ will clearly be still excluded.

An attractive and simple model to explore more quantitatively this problem is to take an ellipsoidal Fermi line, corresponding to an ellipsoidal electronic dispersion relation. On the other hand we will keep a simple BCS type attractive interaction, with a constant matrix element which does not vary over the Fermi surface. Clearly this is not the most general situation, since we might consider a variation of this coupling on the Fermi surface. This would naturally give additional degrees of freedom to the problem, which could combine in various ways to the results we will find. But it is reasonable to investigate first the simplest case, where the coupling is constant.

Surprisingly we find that such an elliptical dispersion relation does not lift the degeneracy found for the isotropic case and that the cascade of phase transitions is not modified. Although the reason for this will appear to be rather simple, this is a somewhat unexpected result which is of quite important practical interest. Indeed we mentionned above that it is not always the case that a circular Fermi line is a bad approximation for a real compound. It is clear that the Fermi lines of many more compounds can be closely approximated by an ellipse. Hence in order to have a model which displays generic features resulting from the anisotropy of the dispersion relation, we include a small additional term to the ellipsoidal dispersion relation which we consider as a perturbation and treat to first order. This corresponds to the fairly large class of compounds where the Fermi line is not so far from being elliptical. This leads us to the expected situation where, except for discrete symmetries, a single wavevector is favored just at the transition. However we will show explicitely that, in agreement with the analysis given above, the cascade of transition appears when one enters the superfluide phase.
Hence this cascade is not destroyed by anisotropy, as one might have feared, but it is merely modified in the way one could expect qualitatively.

The paper is organized as follows. In the next section we consider the second order term in the Landau-Ginzburg expansion, which is the one for which anisotropy is expected to give the most important modification. After handling in the first subsection the case of the purely elliptical dispersion relation, the more complex case of a small deviation from this ideal situation is treated in the second subsection. Then the fourth order term in the Landau-Ginzburg expansion, necessary to obtain the cascade of transitions, is investigated in the third section and it is shown that one may simply use the results already obtained in the isotropic case. Finally in the fourth section we study the resulting modification of the cascade.

\section{SECOND ORDER TERM}
As indicated in the preceding section we take the simplest model for the electronic interaction responsible for pairing, that is a contact potential $V({\bf r}) = V \delta ({\bf r})$, since our purpose is  to concentrate on the effect of the anisotropy in the electronic dispersion relation. This potential is just formally as in the standard BCS theory. Since we expect our transitions to be second order, we will use a Ginzburg-Landau approach and expand the free energy difference $\Omega \equiv  
\Omega _{s} - \Omega _{n}$ between the superconducting and the 
normal state up to fourth order in the order parameter. Hence we write $\Omega =  
\Omega _{2} + \Omega _{4}$ where $\Omega _{2} $ and $\Omega _{4} $ are respectively the second order and the fourth order term. Our main problem is to calculate the second order term in the presence of anisotropy. In terms of the Fourier components $\Delta _{{\bf  q}}$ of 
the order parameter $\Delta ({\bf  r})$: 
\begin{eqnarray}
\Delta ({\bf  r}) = \sum _{{\bf  q}_{i}} \Delta _{{\bf  q}_{i}} 
\exp(i{\bf  q}_{i}. {\bf  r})
\label{eqfourdelt}
\end{eqnarray}
it reads:
\begin{eqnarray}
\Omega_{2} = \sum_{{\bf  q}} \Omega _{2}(q, \bar{ \mu 
},T)
 | \Delta _{{\bf  q}} |^{2}
\label{2freeener}
\end{eqnarray}
The general expression of $\Omega _{2}({\bf q}, \bar{ \mu 
},T)$ is \cite{rccm} (with slightly changed notation): 
\begin{eqnarray}
\Omega _{2}({\bf q}, \bar{ \mu },T) =  \frac{1}{V} - 
T  \sum_{n,k}  \bar{G}({\bf k}) G({\bf k}+{\bf q})
\label{eq2ord}
\end{eqnarray}
where the up spin free electron propagator is given by $ G( {\bf k}) = (i \bar{\omega  } _{n} - \xi _{{\bf  k}}) ^{-1}$ and $ \bar{G}( {\bf k}) = (-i \bar{\omega  } _{n}  - \xi _{{\bf  k}}) 
^{-1}$. Here $\bar{\omega  }_{n} \equiv \omega _{n} - i \bar{\mu }$, 
with  $ \omega _{n} = \pi T (2n+1) $ being the Matsubara frequency and $ \bar{ \mu } = (\mu _{\uparrow} - \mu_{\downarrow})/2 $ 
is half the chemical potential difference between up spin and down spins. Moreover $\xi _{ {\bf  k}} = \epsilon_{{\bf k}} - \epsilon _F$ is the kinetic energy 
measured from the Fermi surface for $ \bar{ \mu } = 0 $ and $\epsilon _F$ is the corresponding Fermi energy.

It is again convenient to introduce the corresponding expression for ${\bf q}={\bf 0}$:
\begin{eqnarray}
A_{0}( \bar{ \mu },T) =  \frac{1}{V} - 
T  \sum_{n,k}  \bar{G}({\bf k}) G({\bf k})
\label{eqa0}
\end{eqnarray}
For $ \bar{ \mu } = 0$, the standard critical temperature $T_{c0}$ of the BCS phase corresponds to $A_{0}( 0,T_{c0}) = 0$ and more generally $A_{0}( \bar{ \mu } ,T_{sp}) = 0$ gives as a function of 
$ \bar{ \mu } $ the temperature $ T_{sp}(\bar{ \mu })$ of the spinodal transition line, where the normal state becomes absolutely unstable against the transition toward the standard uniform BCS phase. Above the tricritical point this line is identical to the standard second order transition toward the BCS phase, but below this tricritical point it is located inside the domain for the superfluid phase since it is superseded by the FFLO transition. Hence we obtain:
\begin{eqnarray}
\Omega _{2}({\bf q}, \bar{ \mu },T) =  A_{0}( \bar{ \mu },T) + I ({\bf q}, \bar{ \mu },T)
\label{eqgdi}
\end{eqnarray}
with:
\begin{eqnarray}
 I ({\bf q}, \bar{ \mu },T) = - T  \sum_{n,k}  [ \bar{G}({\bf k}) G({\bf k}+{\bf q}) -  \bar{G}({\bf k}) G({\bf k}) ]
\label{eq2ord1}
\end{eqnarray}
where the summation over the Matsubara frequencies can be extended to infinity, without any need for the standard BCS cut-off, since this summation is fully convergent. As already indicated in Ref. \cite{rccm}, $A_{0}( \bar{ \mu },T)$ is an increasing function of $\bar{ \mu }$. Since we have $\Omega _{2}=0$ at the second order phase transition, we see from Eq.(\ref{eqgdi}) that the highest value  of $\bar{ \mu }$ for which this transition will appear corresponds to the minimum of value of $I ({\bf q}, \bar{ \mu },T)$.

\subsection{Elliptical Fermi surface}
We consider first an elliptical dispersion relation:
\begin{eqnarray}
\epsilon_{{\bf k}} = \frac{k_{x}^{2}}{2m_{x}} + \frac{k_{y}^{2}}{2m_{y}}
\label{eqdisp}
\end{eqnarray}
for which all the surfaces of equal energies, and in particular the Fermi surface (or precisely the Fermi line, since we work in two spatial dimensions), are ellipses. We make no specific assumptions on the effective masses $m_x$ and $m_y$, so the anisotropy can be very strong. It would seem that this should strongly modify the transition, and in particular that the anisotropy will select specific wavevectors ${\bf q}$ for the space variation of the order parameter. One would expect that one of the major axes of the Fermi surface ellipse is preferred. It is easy to see that this is not at all the case.

Indeed we can convert the summation over ${\bf k}$ in Eq.(\ref{eq2ord1}) into an integral, and then make the change of variables $k_x = \sqrt{m_x} K_x$ and $k_y = \sqrt{m_y} K_y$, which gives $dk_x dk_y = \sqrt{m_x m_y} K_x K_y$, and transforms the electronic dispersion relation into $2 \epsilon _k = K_{x}^{2} + K_{y}^{2} = K^{2}$. If we perform the same change for ${\bf q}$, i.e. we set $q_x = \sqrt{m_x} Q_x$ and $q_y = \sqrt{m_y} Q_y$, we have $ 2\epsilon_{{\bf k}+{\bf q}} = ({\bf K}+{\bf Q}) ^{2}$. Hence the integral we find after these changes is formally just the same as for an isotropic dispersion relation (except that we have to set the mass $m=1$) and we can take up the results of Ref.\cite{rccm}. This will be seen again in the next section when we will calculate this integral in another way. For example this gives at $T=0$: 
\begin{eqnarray}
 I ({\bf q}, \bar{ \mu },0)  = \frac{\sqrt{m_x m_y}}{2\pi }[  {\rm Re} \ln ( 1+ \sqrt{1-\bar{Q}^{2}}) - \ln 2 ]
\label{eq2ord2}
\end{eqnarray}
where we have set $\bar{Q}=(Q/\bar{\mu }) \sqrt{\epsilon_F/2 }$. We obtain that the minimum of $I ({\bf q}, \bar{ \mu },0) $ is $-(\ln 2/2\pi ) \sqrt{m_x m_y} $, which is reached for $\bar{Q}=1$, that is $\epsilon _{\bf q}=\bar{\mu }^{2}/\epsilon _F$. Therefore the modulus $q$ of the wavevector corresponding to the minimum depends on the direction of ${\bf q}$, but the minimum of $I ({\bf q}, \bar{ \mu },0) $ itself is the same whatever this direction : we have a complete degeneracy with respect to the direction of this wavevector, just as in the isotropic case. Naturally this conclusion is also valid at non zero temperature. Accordingly we reach the surprising conclusion that the cascade of transitions which has been found for the isotropic situation \cite{rccm} is not modified when we have an elliptical dispersion relation (this will be fully justified below when we will consider the fourth order term). 

\subsection{Nearly elliptical Fermi surface}

Since we have just seen that an elliptical dispersion relation does not lift the degeneracy with respect to the direction of ${\bf q}$, we have to go away from this simple dispersion relation and consider a more general case. In order to integrate Eq.(\ref{eq2ord1}) over ${\bf k}$, we will use the fact that the dominant contribution arises actually from the vicinity of the Fermi line. This allows to integrate over the variable $\xi_{\bf k}$ and to be left with an integration on the curvilinear abscissa $s_{k}$ along the Fermi line $S_F$. One obtains:
\begin{eqnarray}
 \sum_{k}  \bar{G}({\bf k}) G({\bf k}+{\bf q}) = \pi \; \rm{sign} \, \omega _{n} \frac{1}{(2\pi )^{2}} \int_{S_F} ds_{k} \, \left|{\bf \nabla}.\epsilon _{k}\right|^{-1} \; \frac{1}{\bar{\omega  } _{n} + \frac{\mathit{i}}{2} {\bf q}.{\bf \nabla} \epsilon _{k}}
\label{eqGG1}
\end{eqnarray}
where the inverse modulus $\left|{\bf \nabla}.\epsilon _{k}\right|^{-1}$ of the gradient of $\epsilon _{k}$  is essentially the local density of states.

Actually we will not consider this general situation, but only the case where the departure from the elliptical dispersion relation is small in order to be able to perform explicitely the last integration. We could make a Fourier expansion of this small departure along the Fermi line. But in the generic situation the result will be dominated by the lowest order relevant Fourier component and we will restrict ourselves to the case where only this component is present. Since we will be concerned only by the wavevectors which are near those minimizing our second order term $\Omega _2$, this is enough to obtain fairly general conclusions. Therefore this leads us to consider the dispersion relation:
\begin{eqnarray}
\epsilon_{{\bf k}} = \frac{k_{x}^{2}}{2m_{x}} + \frac{k_{y}^{2}}{2m_{y}} + \frac{2 \eta }{m_x m_y \epsilon _F} \; k_{x}^{2} k_{y}^{2}
\label{eqdisp1}
\end{eqnarray}
where $\eta \ll 1$. Depending on the sign of $\eta$, this corresponds to a Fermi line which is, with respect to an ellipse, slightly deformed toward a rectangle or toward a rhombus. Lower order Fourier components would only shift the ellipse or change its axes. 

It is then simpler to handle this case directly, rather than making use of the above general formula. We make the change of variables from ($k_x$ , $k_y$) to ($\rho$ , $\varphi$) defined by:
\begin{eqnarray}
k_x = \sqrt{m_x}\; \rho \;\cos \varphi \;[ 1 - \frac{\eta}{\epsilon _F}  \rho^{2} \sin ^{2} \varphi ] \nonumber \\
k_y = \sqrt{m_y} \;\rho \;\sin \varphi \;[ 1 - \frac{\eta}{\epsilon _F}  \rho^{2} \cos ^{2} \varphi ]
\label{eqchvar}
\end{eqnarray}
When this is substituted in Eq.(\ref{eqdisp1}), this gives $\epsilon_k = \rho^{2}/2$ to first order in $\eta$. We then make directly this change of variable in Eq.(\ref{eq2ord1}). The Jacobian is found to be:
\begin{eqnarray}
\left|\frac{\partial(k_x,k_y)}{\partial(\rho,\varphi)}\right| = \sqrt{m_x m_y} \,\rho \; (1-\frac{\eta}{\epsilon _F}\rho^{2})
\label{eqjacob}
\end{eqnarray}
Since we have $\epsilon_k = \rho^{2}/2$, we can again integrate over $\xi_k = \epsilon _k - \epsilon _F $. This leads to:
\begin{eqnarray}
 \sum_{k}  \bar{G}({\bf k}) G({\bf k}+{\bf q}) = \pi \; \sqrt{m_x m_y} \; (1-2\eta )\; \rm{sign} \, \omega _{n}  \,\frac{1}{(2\pi )^{2}} \int_{0}^{2\pi } d\varphi \; \frac{1}{\bar{\omega  } _{n} + \frac{\mathit{i}}{2} {\bf q}.{\bf \nabla} \epsilon _{k}}
\label{eqGG2}
\end{eqnarray}

Although the calculation can be carried on at non zero temperature, we will restrict ourselves to the $T=0$ case in the following. Indeed we want to know how the second order term is modified by a small anisotropy and how the cascade of phase transitions is affected. Since this cascade occurs at low temperature, we can compare the effect of the small anisotropy, at $T=0$, to the effect of a small non zero temperature, evaluated for zero anisotropy. This last effect is already known from preceding work \cite{rccm}. At $T=0$ the summation over Matsubara frequencies occuring in Eq.(\ref{eq2ord1}) can be replaced by an integral, which leads to:
\begin{eqnarray}
 I ({\bf q}, \bar{ \mu },0) =  \sqrt{m_x m_y} \; \frac{1-2\eta}{2\pi } \int_{0}^{2\pi }  \frac{d\varphi }{2\pi } \; \ln \left|1-\frac{1}{2\bar{\mu}} {\bf q}.{\bf \nabla} \epsilon _{k} \right|
\label{eq2ord3}
\end{eqnarray}
This formula is very similar to the corresponding one obtained by LO \cite{larkov}. Actually the factor $(1-2\eta )$ can be understood as an effective modification of the density of states. Naturally in Eq.(\ref{eq2ord2}) we have to take $\epsilon _k$ as it is given by the dispersion relation Eq.(\ref{eqdisp1}), which leads explicitely to:
\begin{eqnarray}
\frac{1}{2\bar{\mu}} {\bf q}.{\bf \nabla} \epsilon _{k} = \bar{Q}\, [\cos\varphi + 3\eta \,\sin 2(\varphi +\alpha ) \sin(\varphi+2\alpha )]
\label{eqqgrad}
\end{eqnarray}
where, together with the change Eq.(\ref{eqchvar}), we have set $q_x = \sqrt{m_x} \,Q \cos \alpha $ and $q_y = \sqrt{m_y} \,Q\sin\alpha $, with $\bar{Q}=(Q/\bar{\mu }) \sqrt{\epsilon_F/2 }$, and already made the change of variable $\varphi \rightarrow \varphi + \alpha $.

We are interested in finding the minimum of $ I ({\bf q}, \bar{ \mu },0) $. The situation is here quite analogous to the one found without anisotropy. In this last case $ I ({\bf q}, \bar{ \mu },0) $ is given by Eq.(\ref{eq2ord2}) and its derivative for its minimum, obtained for $\bar{Q}=1$, is discontinuous. It is $-\infty$ for $\bar{Q} \rightarrow 1_{-}$ and is finite and positive for $\bar{Q} \rightarrow 1_{+}$, as it can be seen directly. This is linked to the fact that the argument of the logarithm in Eq.(\ref{eq2ord3}) has a double root for $\bar{Q}=1$. In the presence of anisotropy the behaviour of the argument of the logarithm is similar around its minima and we will find again that the minimum of $ I ({\bf q}, \bar{ \mu },0) $ corresponds also to the case where this argument has a double root $\varphi_{m}$. This implies that this argument and its derivative with respect to $\varphi$ are both zero for $\varphi = \varphi_{m}$. To first order in $\eta$ the derivative is found to be zero for $\varphi_{m} \simeq (9\eta/2) \sin(4\alpha )$. We then request that the argument of the logarithm is zero for $\varphi_{m}$. Actually it can be seen that, up to first order in $\eta$, this is the same as making this argument zero for $\varphi= 0$. This leads to the conclusion that  $ I ({\bf q}, \bar{ \mu },0) $ is minimum for:
\begin{eqnarray}
\bar{Q} = 1 - 3\eta \sin^{2}(2\alpha )
\label{eqqmin}
\end{eqnarray}

We are still left with the evaluation of Eq.(\ref{eq2ord3}) for this value of $\bar{Q}$. It is convenient to take
then the derivative with respect to $\eta$ of the integral in Eq.(\ref{eq2ord3}). It can be seen that this derivative is not singular, so we can evaluate it for $\eta=0$. This can be done analytically and gives merely $-3 \cos(4\alpha )$. Integrating with respect to $\eta$ we finally obtain that, to first order in $\eta$, the minimum of $ I ({\bf q}, \bar{ \mu },0) $ is given by (for a fixed direction of ${\bf q}$, given by $\alpha $):
\begin{eqnarray}
{\rm Min} [ I ({\bf q}, \bar{ \mu },0) ] =  - \sqrt{m_x m_y} \; \frac{1-2\eta}{2\pi }\, \left[  \,\ln 2 + 3 \eta \cos 4\alpha \,\right]
\label{eqimin}
\end{eqnarray}
This result shows that the degeneracy with respect to the direction of ${\bf q}$ is lifted as we expected. The minimum of  $I ({\bf q}, \bar{ \mu },0)$ is found for $\cos 4\alpha =1$, i.e. $\alpha = n \pi /2$, where $n$ is an integer. This corresponds to have the direction of ${\bf q}$ along any axis of our nearly elliptical Fermi line.

\section{FOURTH ORDER TERM}

The fourth order term $\Omega _{4}$ in the free energy difference is given \cite{larkov,rccm} by:
\begin{eqnarray}
\Omega_{4} = \frac{1}{2} \sum_{{\bf  q}_{i}} \Omega _{4}({{\bf  
q}}_{1},{{\bf  q}}_{2},{{\bf  q}}_{3},{{\bf  q}}_{4},\bar{ \mu 
},T) \Delta _{{\bf  q}_{1}} \Delta ^{*} _{{\bf  q}_{2}} \Delta _{{\bf  
q}_{3}}  \Delta ^{*} _{{\bf  q}_{4}}
\label{4freeener}
\end{eqnarray}
with:
\begin{eqnarray}
\Omega _{4}({{\bf  q}}_{1},{{\bf  q}}_{2},{{\bf  
q}}_{3},{{\bf  q}}_{4},
\bar{ \mu },T) =  T  \sum_{n,k}  \bar{G}({\bf k}) G({\bf k}+{\bf 
q}_{1})
\bar{G}({\bf k}+{\bf q}_{1}-{\bf q}_{4}) G({\bf k}+{\bf q}_{2})
\label{eq12}
\end{eqnarray}
and where the ${\bf  q}_{i}$'s satisfy momentum conservation ${\bf  q}_{1} + 
{\bf  q}_{3} = {\bf  q}_{2} + {\bf  q}_{4}$. 
In the case of an elliptical dispersion relation $\epsilon_{{\bf k}} = k_{x}^{2}/2m_{x} + k_{y}^{2}/2m_{y}$
we can make the same change of variables $k_i = \sqrt{m_i} K_i$ as for the second order term, and set again $q_i = \sqrt{m_i} Q_i$. Just as above the integral we find after these changes is formally just the same as for an isotropic dispersion relation (with $m=1$). Hence, as indicated above, the cascade of
phase transitions is completely unaffected when the isotropic dispersion relation is transformed in an
elliptical one.

In order to find a modification we have to go away from an elliptical dispersion relation and consider the dispersion relation Eq.(\ref{eqdisp1}). In order to minimize the free energy for fixed directions of the wavevectors ${\bf  q}_{i}$ we have first to take their lengths in such a way that they minimize the second order term in the free energy, and then to evaluate the fourth order term for these lengths. Since in the second term the contributions of the various wavevectors add up, as it is seen in Eq.(\ref{2freeener}), we have just to minimize the contribution of each wavevector. At zero temperature, for a given wavevector direction $\alpha $, the length corresponding to the minimum is given by Eq.(\ref{eqqmin}). However we have to take into account that, at non zero temperature, this result is modified. In the isotropic case this modification has been obtained in \cite{rccm} as:
\begin{eqnarray}
\bar{q}-1 = \frac{t}{2} \,  \ln  \frac{\pi }{2t} 
\label{leading2}
\end{eqnarray}
where $ t = T /  \bar{\mu}$. As we have seen,  this result is unchanged for an elliptical dispersion relation provided we replace $\bar{q}$ by $\bar{Q}$. We have then to compare the anisotropy modification Eq.(\ref{eqqmin}) to the temperature modification Eq.(\ref{leading2}). This depends on the angle $\alpha $ for the considered wavevector. However we will be interested only by wavevectors with a direction near the one which gives the absolute minimum of the second order term, that is near $\alpha=0 $ (or $\alpha = n \pi /2$). The departure from this angle will be typically the angular difference $\alpha _c$ between two wavevectors appearing in the cascade for the isotropic case, which has been found \cite{rccm} to satisfy $\alpha_{c}^{2}= 8t \ln ((2 \pi ^{3})^{1/2}/t)$ when $t$ is small. Carrying this result into Eq.(\ref{eqqmin}) we see that the temperature dependences of the two modifications are the same, with naturally the contribution Eq.(\ref{leading2}) being dominant in the limit $\eta \rightarrow 0$, although there is a large numerical coefficient. Anyway the temperature dependence of the fourth order term will not be modified by anisotropy. For simplicity we assume in the following the small anisotropy situation $\eta \ll 1$. In this case we are only left with the dominant modification Eq.(\ref{leading2}). Accordingly the fourth order term is not modified by anisotropy, and we can make use of the isotropic results \cite{rccm} .

\section{THE  CASCADE  OF  PHASE  TRANSITIONS}

In this section we will show that, with the anisotropy that we have considered, the cascade of phase transitions found in the isotropic case will also appear, although there are naturally quantitative modifications. Indeed we have now a competition between the fourth order term which favors the appearance of the cascade, and the anisotropy in the second order term which favors wavevectors  direction near the preferred direction. Clearly this makes the quantitative study more complicated since we have no longer any symmetry reason to argue that all the wavevectors in the expansion Eq.(\ref{eqfourdelt}) will have equal weight nor any reason to have them with equal angular spacing. Hence a detailed quantitative study necessarily involves a good deal of numerical analysis. We will not go in such details since the qualitative trend, in which we are actually mostly interested, is on the other hand quite clear.

The general expression Eq.(\ref{4freeener}) for the fourth order term in the free energy can be rewritten explicitely \cite{larkov} as:
\begin{eqnarray}
\Omega_{4} = \frac{1}{2} \sum_{i,j} (2- \delta _{{\bf  Q}_{i},{\bf  Q}_{j}}) |  \Delta 
_{{\bf  Q}_{i}}| ^{2}  |  \Delta _{{\bf  Q}_{j}}| ^{2} J(\alpha _{{\bf  
Q}_{i},{\bf  Q}_{j}}) + (1-\delta _{{\bf  Q}_{i},{\bf  Q}_{j}}-\delta 
_{{\bf  Q}_{i},-{\bf  Q}_{j}})  \Delta  _{{\bf  Q}_{i}}  \Delta _{-{\bf  
Q}_{i}}  \Delta ^{*} _{{\bf  Q}_{j}}  \Delta  ^{*}_{-{\bf  Q}_{j}}  
\tilde{J} (\alpha _{{\bf  Q}_{i},{\bf  Q}_{j}})
\label{fourthorder}
\end{eqnarray}
by making use of the fact that all the four involved wavevectors have same length and using momentum conservation.

The angular dependence of $J(\alpha _{{\bf  Q}_{i},{\bf  Q}_{j}})$ and $ \tilde{J} (\alpha _{{\bf  Q}_{i},{\bf  Q}_{j}})$ has been already studied \cite{rccm}. This is the quite singular behaviour of  $J(\alpha) $ which is responsible for the appearance of the cascade. $\tilde{J} (\alpha)$ is less singular and never plays a dominant role. Hence we omit it from now on for simplicity, since it does not modify the qualitative picture. As it has been shown in Ref. \cite{rccm}, $J(\alpha )$ is quite large positive at low temperature for $\alpha =0$. On the other hand it drops very rapidly and becomes negative for some angle $\alpha _c$, which has already been introduced in the preceding paragraph. The minimum of $J(\alpha )$ is very near $\alpha _c$. Hence one can view the "interaction" of two wavevectors as being strongly repulsive
for small angle, and attractive beyond $\alpha _c$ with the optimum angle being essentially for $\alpha _c$. This is the physical feature which gives rise to the cascade. We will show that it produces the same effect when anisotropy is present. 

For this purpose we will compare the free energy for an order parameter with a single wavevector parallel to the direction $\alpha =0$, favored by the second order term, to the free energy of an order
parameter with three vectors, one of them in the direction $\alpha =0$ and the two other ones in symmetrical directions with angles $\pm \alpha _c$. Actually we omit here an additional complication which is again of secondary importance. Indeed it is always more advantageous \cite{rccm} to have wavevectors by pairs with opposite directions. Naturally this is neutral with respect to the second order term, which is unchanged when the direction of the wavevector is reversed. On the other hand it is easily seen that this is energetically favorable from the effect of the $J(\alpha )$ term, since $J(\pi ) < 0$. Moreover the $ \tilde{J} (\alpha )$ that we are omitting favors also this geometry. As a result the order parameter is a combination of cosines, and for example an order parameter with a single plane wave does not occur. However since $|J(\alpha )|$ is fairly small for $\alpha =\pi $, this brings in practice little change in the free energy evaluation, and we omit this complication for simplicity.

Naturally since we need the effect of the anisotropy to be balanced by the fourth order term, the three wavevector phase will appear only when we penetrate in the superfluid domain, but not far from the normal to superfluid transition since anisotropy is small. We evaluate first the second order term under these conditions. It is still given by Eq.(\ref{eqgdi}). The expression of $I ({\bf q}, \bar{ \mu },0)$ depends only on the ratio ${\bf q}/ \bar{ \mu }$ and has already been treated in section II. We have only to take into account the fact that $A_{0}( \bar{ \mu },T)$, given by Eq.(\ref{eqa0}), depends on the location of ($T,\bar{ \mu }$) with respect to the transition line. In its vicinity it will vary linearly with the parameters $T$ and $ \bar{ \mu }$. Since we are at low temperature it will depend essentially only on $\bar{ \mu } $. Taking the derivative of Eq.(\ref{eqGG2}) with respect to $\bar{ \mu } $, in the case ${\bf q}=0$, we obtain:
\begin{eqnarray}
\Omega _{2}({\bf q}, \bar{ \mu },0) = N_0 \left[\frac{\bar{ \mu } }{\bar{ \mu }_c }-1 + 3\,\eta (1-\cos 4\alpha)\right]
\label{eqa0mu}
\end{eqnarray}
where the evaluation has been made for the optimum value of $q$ for a given angle $\alpha $. We have taken Eq.(\ref{eqimin}) into account and we have set $N_0 =\sqrt{m_x m_y} \; ((1-2\eta)/2\pi )$, which is the straightforward generalization of the isotropic result. Here ${\bar{ \mu }_c}$ is the critical effective field for the normal to (FFLO) superfluid phase at $T=0$, for $\alpha =0$.

Calling respectively $|\Delta _0|^{2}$ and $|\Delta _1|^{2}$ the weights of the order parameter $|\Delta _{{\bf q}}|^{2}$ for $\alpha =0$ and $\alpha = \alpha _c$, we have for the free energy of the three wavevector state:
\begin{eqnarray}
\frac{\Omega}{N_0} = {\bar \Omega _{0}} |\Delta _0|^{2} + 2{\bar \Omega _{1}} |\Delta _1|^{2} + \frac{1}{2} J_0 (|\Delta _0|^{4} + 2 |\Delta _1|^{4}) + 4 J_1 |\Delta _0|^{2} |\Delta _1|^{2} + 2 J_2 |\Delta _1|^{4}
\label{eqfreen3}
\end{eqnarray}
where we have set ${\bar \Omega _{0}}=\Omega _{2}(0, \bar{ \mu },0) $, ${\bar \Omega _{1}}=\Omega _{2}(\alpha _c, \bar{ \mu },0) $, $J_0=J(0)$,$J_1=J(\alpha _c)$ and $J_2=J(2 \alpha _c)$. The minimization with respect to $|\Delta _0|^{2}$ and $|\Delta _1|^{2}$ leads to:
\begin{eqnarray}
{\bar \Omega _{0}} + J_0 |\Delta _0|^{2} + 4 J_1 |\Delta _1|^{2} = 0 \\
2{\bar \Omega _{1}} + (2 J_0 +4 J_2)|\Delta _1|^{2} + 4 J_1 |\Delta _0|^{2} = 0
\label{eqminimi}
\end{eqnarray}
In practice we can simplify these equations because, as we have seen \cite{rccm} , we have at low temperature $|J_1| \ll J_0$ and $|J_2| \ll J_0$. This leads merely to $|\Delta _0|^{2}=-{\bar \Omega _{0}}/J_0$ and $|\Delta _1|^{2}=-{\bar \Omega _{1}}/J_0$, leading to the free energy:
\begin{eqnarray}
\frac{\Omega}{N_0} = -\frac{{\bar \Omega _{0}}^{2}+2{\bar \Omega _{1}}^{2}}{2J_0} 
\label{eqfreenmin}
\end{eqnarray}
This is to be compared to the free energy $\Omega/N_0 = -{\bar \Omega _{0}}^{2}/2J_0$ for a single wavevector with $\alpha =0$, obtained simply by setting $|\Delta _1|^{2} =0$ from the start. This simple result Eq.(\ref{eqfreenmin}) means merely that the dominant effect is that one gains free energy by increasing the number of wavevectors, because one lowers comparatively the effect of the self repulsive term $J_0$ in the fourth order term. This is the main origin \cite{rccm} of the cascade.

Naturally this occurs only when $|\Delta _1| \neq 0$ can be satisfied, that is when $\bar \Omega _{1} <0$. This means that the second order term is now negative not only for $\alpha =0$, but also for $\alpha = \alpha _c$. From Eq.(\ref{eqa0mu}) this gives explicitely:
\begin{eqnarray}
1 - \frac{\bar{ \mu } }{\bar{ \mu} _c } > 3\,\eta \,(1-\cos 4\alpha_c)
\label{eqfin}
\end{eqnarray}
For a small departure from an ellipsoidal dispersion relation, i.e. $\eta \ll 1$ this implies that one has to make only a small penetration into the superfluid domain to meet the three wavevectors state. Since at low temperature $\alpha _c$ is small and we have its asymptotic behaviour \cite{rccm} given above in this regime, we have more explicitely at low temperature for the boundary of the domain where the three wavevector state can be found:
\begin{eqnarray}
1 - \frac{\bar{ \mu } }{\bar{ \mu} _c } > 24 \, \eta  \alpha_c^{2} \sim 192 \,\eta \, t \ln ((2 \pi ^{3})^{1/2}/t)
\label{eqfin1}
\end{eqnarray}
Although the numerical coefficient is large, one has to keep in mind that the right-hand side of this inequality is always small since it is necessarily much less than $3\,\eta$. One can also see qualitatively how this result is generalized to states with more than three wavevectors. For example the state with five wavevectors will appear when the second order term becomes negative for the angle $2 \alpha _c$. For small anisotropy, this is again reached for a small penetration into the superfluid domain. One can also investigate along the same lines when it is energetically favorable to have two wavevector state with angles $\pm \alpha _c/2$. With the same simplifications as above, the free energy of this state is found to be $-{\bar \Omega _{1/2}}^{2}/J_0$ where ${\bar \Omega _{1/2}}=\Omega _{2}(\alpha _c/2, \bar{ \mu },0) $. By comparing it to the free energy $-{\bar \Omega _{0}}^{2}/2J_0$ for the single wavevector with $\alpha =0$ we see that the two wavevectors state arises when $\sqrt{2} {\bar \Omega _{1/2}} > {\bar \Omega _{0}}$, that is:
\begin{eqnarray}
1 - \frac{\bar{ \mu } }{\bar{ \mu} _c } > \frac{\sqrt{2}}{\sqrt{2}-1}\,3\,\eta \,(1-\cos 2\alpha_c) \simeq 20.5  \, \eta  \alpha_c^{2}
\label{eqfin2}
\end{eqnarray}
Hence this state will appear slightly before the three wavevectors state when one enters into the superfluid domain of the phase diagram.

\section{CONCLUSION}

In this paper we have studied the transition from the normal state to the FFLO phases in a two dimensional system with anisotropy. More specifically we have considered how anisotropy modifies the cascade of phase transitions, found for an isotropic system when the temperature is lowered. We have first considered a simple model with an ellipsoidal Fermi line, corresponding to an elliptical dispersion relation. We have found surprisingly that such a modification of the dispersion relation does not remove the degeneracy of all wavevectors ${\bf q}$ on the Fermi line with respect to the appearance of a FFLO phase. More generally we have found that, in this case, the cascade of phase transitions obtained in the isotropic case is not changed. We have then considered how these results are modified when a small term is added to the ellipsoidal dispersion relation, in order to eliminate this degeneracy. We have shown that the degeneracy is indeed removed and that, essentially, a single wavevector is selected right at the transition. However when one enters in the superfluid phase, a cascade of phase transitions still appears. This shows that, quite generically, the cascade found for the isotropic case is modified, but not destroyed, by anisotropy.

We are grateful to A. Buzdin and C. Mora for stimulating discussions.

* Laboratoire associ\'e au Centre National
de la Recherche Scientifique et aux Universit\'es Paris 6 et Paris 7.

\end{document}